

\input amstex.tex
\documentstyle{amsppt}
\magnification\magstep1
\topmatter\title Vacuum expectation values of products of\\ chiral currents in
$3+1$
dimensions\endtitle
\author Jouko Mickelsson\\May 31, 1992 \endauthor
\affil Center for Theoretical Physics,
M.I.T., Cambridge, MA 02139, USA. Permanent address:
Department of Mathematics, University of Jyv\"askyl\"a, SF-40100,
Jyv\"askyl\"a, Finland\endaffil
\endtopmatter

\define\a{\alpha}
\redefine\b{\beta}

\define\e{\epsilon}
\define\G{\Gamma}
\redefine\o{\omega}
\redefine\O{\Omega}
\redefine\l{\lambda}
\redefine\L{\Lambda}
\define\RM{\Bbb R}
\define\CM{\Bbb C}

\define\gm{\bold g}
\define\hm{\bold h}

\define\<#1,#2>{\langle #1,#2\rangle}
\define\TR{\text{tr}}
\define\dep(#1,#2){\text{det}_{#1}#2}
\define\norm(#1,#2){\parallel #1\parallel_{#2}}

\document
\NoBlackBoxes

ABSTRACT An algebraic rule is presented for computing expectation values of
products of local nonabelian charge operators for fermions coupled to an
external vector potential in $3+1$ space-time dimensions. The vacuum
expectation
value of a product of four operators is closely related to a cyclic cocycle
in noncommutative geometry of Alain Connes. The relevant
representation of the current is constructed using Kirillov's method of
coadjoint orbits.

\vskip 0.4in
\bf 1. Introduction \rm

\vskip 0.3in
In $1+1$ space-time dimensions it is known that a normal ordering of local
charge operators is sufficient to make them well-defined in a suitable
dense domain of a fermionic Fock space. Assuming that the physical space is
compactified as the circle $S^1,$ the normal ordered charge densities define
a representation of a central extension $\widehat{LG}$ of the (Lie algebra of)
the loop group $LG$ corresponding to a compact gauge group $G.$ In the case of
chiral fermions the central term is nontrivial and gives rise to a highest
weight representation of $\widehat{LG}.$ The Lie algebra of the group is an
affine Kac-Moody algebra.

In higher dimensions even the normal ordered current densities do not give
well-defined operators in the Fock space. Even the state created from the
vacuum by an action of a typical element of the current algebra has an
infinite norm. This reflects the fact that the automorphisms of the algebra
of canonical anticommutation relations (CAR) generated by gauge transformations
are not implementable by unitary transformations in the Fock space in space
dimensions higher than one. (For a thorough discussion of CAR representations
see [A].) In other words, a gauge transformation tends to
take a state in the Fock space to a vector in a different Fock space
corresponding to an \it inequivalent \rm representation of CAR.

One is thus lead to consider, not a representation in a single Fock space,
but an action of the group of gauge transformations in a bundle of Fock spaces,
parametrized by vector potentials. The representations of CAR in different
fiber
s
are in general nonequivalent.

In any case, in quantum field theory we would like to compute matrix elements
like
$$(\phi, X_1X_2\dots X_n\psi),\tag1.1$$
where $\phi,\psi$ are states in the quantum Hilbert space and the $X_i'$s are
smeared charge operators, formally
$$X_i=\int_M J_0(x) f(x) dx,\tag1.2$$
where $f$ is a test function, $J_0$ is the charge density and the integral is
taken over the three dimensional physical space $M.$ But how can we make
sense of (1.1) if
the charges are not well-defined operators in the Fock space? The answer is
that the $X_i$'s still make sense as sesquilinear forms in the Fock space, [R],
and therefore expressions like $(\phi,X\psi)$ are meaningful when the arguments
$\phi,\psi$ are restricted to a suitable dense domain $\Cal F_0\subset\Cal F$
in
the Fock space. But what about expressions like $(\phi,X_1X_2\psi)$ involving
products of charge operators? A sesquilinear form can be thought of as an
infinite matrix but the usual product of a pair of matrices does not
necessarily
converge in infinite dimensions.

The convergence problem was solved by E. Langmann by introducing a regularized
multiplication of matrices (sesquilinear forms), [L]. Instead of explaining the
details of Langmanns construction I shall give a derivation of the rules
how to compute the expectation values (1.1) which is based on an earlier work
on gauge group actions in Fock bundles, [M]. It will turn out that all the
matrix elements of operator products $X_1X_2\dots X_n$ can be evaluated using
simple algebraic relations based on Lie algebra extensions generalizing the
structure of an affine Lie algebra to higher dimensions.

In order to get to the bare essentials of the construction it is useful to
consider a bigger group than the group $Map(M,G)$ of smooth gauge
transformations. To make things technically slightly simpler we
assume that the physical space $M$ is a compact oriented spin manifold and that
the classical spinor fields are sections of $Spin\otimes E$, where $Spin$ is
the spin bundle (with fiber $\Bbb C^2$) and $E=M\times V$ is a trivial complex
vector bundle over $M$ with a unitary representation $\rho$ of $G$ in the
fiber $V.$ (The compactness of $M$ can be traded off to finiteness of a certain
Sobolev norm of $g-1$, $g\in Map(M,G)$, and the triviality of $E$ can be
dropped
if $Map(M,G)$ is replaced by the group of bundle automorphisms of a principal
bundle $P$ to which $E$ is an associated bundle.)

Let $H$ be the space of square-integrable classical spinor
fields on $M$ (sections of $Spin\otimes E$) and let $H_+$ be the subspace of
$H$ spanned by vectors corresponding to the nonnegative part of the spectrum
of a Dirac operator $D,$ and $H_-$ the orthogonal complement of $H_+.$
Any linear operator $g$ in $H$ can be written as a block matrix
$$g=\left(\matrix a&b\\c&d\endmatrix\right)\tag1.3$$
with respect to the splitting $H=H_+\oplus H_-.$  We denote by $GL_p$ the
group consisting of invertible bounded linear operators in $H$ such that the
off-diagonal blocks $b,c$ are in the Schatten ideal $L_{2p}$ of operators $T$
such that $(T^*T)^p$ has a converging trace.

It can be shown that $Map(M,G)\subset GL_2$ when dim$M=3,$ [MR].
The embedding is given
by the natural action of gauge transformations on spinor fields, that is, by
point-wise multiplication $(T(g)\psi)(x)=\rho(g(x))\psi(x).$
We shall show how to compute matrix elements of products $\hat X_1\dots \hat
X_n
$
where the $X_i$'s belong to the Lie algebra of $GL_2$ and the hat means the
corresponding second quantized operator. By restriction to the Lie algebra of
the subgroup $Map(M,G)$ this will give all the required matrix elements of
local
nonabelian charges.

The quantum sesquilinear forms $\hat X$ form a Lie algebra $\widehat{\bold{gl}
}_2$ which was derived in [MR] from a $\widehat{GL}_2$ action in a (dual)
determinant bundle $DET_2^*$ over a Grassmannian $Gr_2;$ the latter is a
homogeneous space
for $GL_2.$ Each element $W\in Gr_2$ is a closed subspace of $H$ and thus
defines a polarization $H=W\oplus W^{\perp}.$ On the other hand, each
polarizati
on
of $H$ defines a representation of CAR in a Fock space $\Cal F_W$ with a Dirac
vacuum $|W>$ characterized by
$$a^*(u)|W>=0=a(v)|W>\phantom{AAA}\text{for $u\in W^{\perp}$, $v\in
W$}\tag1.4$$
where $a^*(u)$ is a creation operator and $a(v)$ is an annihilation operator
with the only nonvanishing anticommutators
$$a^*(u)a(v)+a(v)a^*(u)=(u,v)\phantom{aaa} u,v\in H.\tag1.5$$

The inner product $(\cdot,\cdot)$ in $H$ is linear in the first and antilinear
in the second argument.
In the case of chiral fermions this construction is "twisted" in such a way
that the bundle $Vac$ of Dirac vacua is equal to the determinant bundle
$DET_2^*,$ [M1]. For this reason the anomalies related to chiral gauge
transform
ations
are given in terms of gauge action in the line bundle $DET_2^*.$ The Schwinger
terms are equal to those found in [MR] and were later derived in [L] using
the sesquilinear approach of Ruijsenaars. (In fact, there was a perturbative
argument [JJ] that certain commutator anomalies would arise when quantizing
a nonabelian chiral Yang-Mills-Dirac system. A geometric and mathematically
consistent treatment was given in [M3,FS].)  On the other hand, it has been
proven by D. Pickrell that "normal" representations of $\widehat{GL}_2$ do
not exist, [P1].

In sections 2 and 3 we shall recall some basic facts about determinant bundles
over Grassmannians and group extensions $\widehat{GL}_p;$ for more details
see [MR], [M2], [PS]. The results on vacuum expectation values are contained
in section 4. In section 5 we reconstruct the $\widehat{GL}_2$ action on
vacuum line bundle from Kirillov's theory of coadjoint orbits. The appropriate
orbit is the cotangent bundle $T^*Gr_2$ and the prequantization line bundle
for Kostant-Souriou quantization is the determinant bundle $DET_2^*$ when
restricted to the zero section in $T^* Gr_2.$

\vskip 0.3in
\bf 2. Basic notions about infinite-dimensional Grassmannians \rm

\vskip 0.3in
Let $H=H_+\oplus H_-$ be a polarization of an infinite-dimensional complex
separable Hilbert space to a pair of closed infinite-dimensional subspaces.
For any closed subspace $W\subset H$ and positive integer $p$ we denote
by $Gr_p(W)$ the Grassmannian consisting of closed subspaces $W'\subset H$
such that
\roster \item the orthogonal projection $W'\to W$ is a Fredholm operator
\item the projection $W'\to W^{\perp}$ is in the Schatten ideal $L_{2p},$
\endroster
We shall denote also $Gr_p=Gr_p(H_+).$ The Grassmannian splits to connected
components $Gr_p^{(k)}(W)$ according to the Fredholm index $k$ of the
projection $W'\to W.$

We shall fix an orthonormal basis $\{e_n\}_{n\in\Bbb Z}$ of $H$ such that
$e_n\in H_+$ for $n\geq 0$ and $e_n\in H_-$ for $n<0.$ If $W\in Gr_p$ then
it has a basis $\{w_n\}_{n>0}$ (not necessarily orthonormal) such that
$$w_n=\sum_{m\geq -k} \a_{mn} e_m + \sum_{m< k} \b_{mn} e_m\tag2.1$$
with $\a-1\in L_1,$ where $k$ is the Fredholm index of the projection
$W\to H_+.$ (In fact, it is possible to choose $w$ such that $\a-1$ is of
finite rank.) We shall call such a $w$ \it an admissible basis \rm and the
set of all admissible basis is a Stiefel manifold $St_p.$

The Stiefel manifold $St_p$ splits to connected components $St_p^{(k)}$
labelled by the Fredholm index $k.$

It is often convenient to think of points  $W\in Gr_p$ as operators $F:H\to H.$
Namely, to each $W$ we can associate the operator $F$ such that $F\vert_{W}=+1$
and the restriction of $F$ to the orthogonal complement of $W$ is $-1.$
Clearly $F^2=1$ and $F^*=F.$ Furthermore, if we write
$$F=\left(\matrix F_{11}& F_{12}\\F_{21}& F_{22}\endmatrix\right)\tag2.2$$
with respect to the splitting $H=H_+\oplus H_-$ then the off-diagonal blocks
are in $L_{2p}$, $F_{11}-1\in L_p$ and $F_{22}+1\in L_p.$ In particular, when
$W=H_+$ then the corresponding operator $\epsilon=F(H_+)$ is
$$\epsilon=\left(\matrix 1&0\\0&-1\endmatrix\right).$$

Differentiating the equation $F^2=1$ we observe that the tangent space to the
Grassmannian at $F\in Gr_p$ is represented by hermitean operators $u$ which
anticommute with $F$ and such that the diagonal blocks of $u$ are in $L_p$ and
the off-diagonal blocks in $L_{2p}.$ By H\"older inequalities, the cotangent
space $T^*_F Gr_p$ consists of hermitean operators $P$ anticommuting with $F$
and with diagonal blocks in $L_{p/(p-1)},$ off-diagonal blocks in
$L_{2p/(2p-1)}
$ The value of the linear form $P$ at $u$ is $\TR uP.$

We also need a generalization of the above definitions. Instead of $H_+$ and
the
basis $\{e_n\}$ we could choose some plane $W\in Gr_p$ with an admissible
basis $w=\{w_1,
w_2,\dots\}$, complete this to a basis $\{w\}_{n\in\Bbb Z}$ of $H$, and define
a basis $w'$ of some $W'\in Gr_{p'}(W)$ to be admissible relative to $w$ if the
matrix $w'(w)$ giving the projection of $w'$ to the vectors $\{w_i\}_{i\geq
-k}$
is of the type
$1+L_1,$ where $k$ is the Fredholm index of the projection $W'\to W.$
We shall denote the Stiefel manifold consisting of these basis $w'$
by $St_{p}(W).$ It does not depend on the choice of the admissible basis $w$ of
$W.$

Let $GL_p$ be as in the introduction. Note that  automatically
the diagonal blocks of $g\in GL_p$ are Fredholm operators with opposite
Fredholm
indices.  The Lie algebra  \define\gl#1{\widehat{\bold{gl}}_#1}
$\bold{gl}_p$ consists of bounded operators in $H$ such that the off-diagonal
blocks are in $L_{2p}.$
The group $GL_p$ acts naturally on $Gr_p.$ In fact, $Gr_p=GL_p/B_p,$ where
$B_p$ consists of the upper triangular matrices, $c=0.$

Let $GL^p$ denote the group of operators $t$ (in some fixed Hilbert space)
such that $t-1\in L_p.$
There is a natural complex line bundle, the determinant bundle $DET_p$, over
$Gr_p.$ Its fiber at $W$ consists of pairs $(w,\l)\in St_p\times \Bbb C$, $w$
being a basis of $W,$ with the equivalence
$$(wt,\l)\sim (w,\l\text{det}\,t), \text{ for }t\in GL^1\tag2.3$$
A section of the dual determinant bundle $DET_p^*$ is then a function
$\psi:St_p\to\Bbb C$ such that $\psi(wt)=\psi(w)\cdot\text{det}\,t$ for
$t\in GL^1.$

\vskip 0.3in
\bf 3. Some group actions on bundles over Grassmannians\rm

\define\GL#1{\widehat{GL}_#1}
\vskip 0.3in
There is an extension $\GL p$ of $GL_p$ by the abelian ideal $Map(Gr_p,\Bbb C^
{\times})$ ($\Bbb C^{\times}$ is the multiplicative group of nonzero complex
numbers) which acts in the total space of the bundle $DET_p.$ The structure
of this group is explained in [MR], [M2]; here we shall recall some basic
facts (in slightly different way than in the references).

Define first the group $\Cal E_p$ consisting of pairs $(g,q)$, where $g\in
GL_p$ and $q$ is a
$\infty\times\infty$-matrix valued function on $Gr_p$ such that $gwq(W)^{-1}$
is an admissible basis of $gW$ for any admissible basis $w$ of $W.$ The
multiplication is defined by
$$(g,q)(g',q')=(gg',q'') \text{ with } q''(W)=q(g'W)q'(W).\tag3.1$$
The  pairs $(1,q)$, $q(W)\in GL^1$  and det$\,q(W)=1,$
for all $W\in Gr_p$ form a normal subgroup. Dividing by this subgroup one
obtains the group $\GL p.$

The action of $\GL p$ in $DET_p$ is given by
$$(g,q)\cdot (w,\l)=(gwq(W)^{-1},\l)\tag3.2$$
where $W$ is the plane spanned by $w.$ The natural action in the space of
sectio
ns
of $DET_p^*$ is
$$[(g,q)\cdot\psi](w)=\psi(g^{-1}wq(W)).\tag3.3$$

The group $\GL p$ is a fiber bundle over $GL_p$ with fiber $Map(Gr_p,\Bbb C^
{\times}).$ Actually, there is a group with smaller fiber which acts in
$DET_p.$
In the case $p=1$ the "regularization" $q$ can be chosen to be a constant
function on $Gr_1$ and we get an extension of $GL_1$ by $\Bbb C^{\times};$ the
structure of this central extension is explained in detail in [PS]. The Lie
algebra $\gl 1$ is a vector space sum $\bold{gl}_1\oplus\Bbb C$ with the
following commutators, [Lu],
$$[(X,\l),(X',\l')]=([X,X'],\frac14\TR\epsilon[\epsilon,X][\epsilon,X']).
\tag3.4$$

When $p=2$ the extension is not central. As a vector space the Lie algebra
$\gl 2$ is a direct sum of $\bold{gl}_2$ and the abelian Lie algebra of maps
$h:Gr_2\to \Bbb C$ of the form
$$h(F)=\a+\TR \xi (\epsilon-F),\tag3.5$$
where $\a\in \Bbb C$ and $\xi:H\to H$ is a linear map such that the
off-diagonal
blocks are in $L_{4/3}$ and the diagonal blocks are in $L_2.$  The commutator
is defined as
$$[(X,h),(X',h')]=([X,X'],\Cal L_X h' -\Cal L_{X'} h
+c_2(X,X';\cdot)),\tag3.6$$
where $\Cal L_X$ denotes the Lie derivative on $Gr_2$ arising from the natural
action of $GL_2$ on the Grassmannian and $c_2$ is the Lie algebra cocycle with
coefficients in the space of functions (3.5), [MR],
$$c_2(X,X';F)=\frac18 \TR(\epsilon-F) [[\epsilon,X],[\epsilon,X']].\tag3.7$$

Following Pickrell [P1] one can think of $\gl 2$ as a central extension of a
Lie  algebra \define\gj{
\bold{gl'}_2} $\gj.$ As a vector space $\gj =\bold{gl}_2\oplus M_{2,4/3},$
where $M_{2,3/4}$ consists of operators $P:H\to H$ such that the diagonal
blocks are in $L_2$ and the off-diagonal blocks are in $L_{4/3}.$ The
commutation relations are
$$[(X,P),(Y,Q)]=([X,Y],[X,Q]-[Y,P]+[[\epsilon,X],[\epsilon,Y]]).\tag3.8$$
Thus without the last term on the right the commutator would define a
semidirect
product of $\bold{gl}_2$ and the abelian algebra $M_{2,4/3}.$

The 2-cocycle defining the central extension $\gl 2=\bold{gl}'_2\oplus\CM$ is
$$\omega((X,P),(Y,Q))=\frac{1}{32}\TR\epsilon[[\epsilon,X],[\epsilon,Q]]
-\frac{1}{32}\TR \epsilon[[\epsilon,Y],[\epsilon,P]].\tag3.9$$
The group $GL'_2$ corresponding to $\gj$ is $GL_2\times M_{2,4/3}$ with the
composition rule
$$\align (g_1,P_1)\cdot (g_2,P_2) &=(g_1g_2,P_1+g_1 P_2 g_1^{-1}+\frac12
[\epsilon,g_1\epsilon g_1^{-1}]\\ &\,\,+\frac12 g_1[\epsilon,g_2\epsilon
g_2^{-1}]g_1^{-1}-\frac12 [\epsilon,g_1g_2\epsilon(g_1g_2)^{-1}]).
\tag3.10 \endalign$$

We shall later need the adjoint action of $\GL 2$ on the Lie algebra $\gl 2.$
Since the center of $\GL 2$ does not contribute to the adjoint action, the
action is determined by the $GL'_2$ action on $\gl 2.$ Using the commutation
relations (3.6) the latter is found to be
$$\align Ad_{(g,P)} (X,Q,\a)&=(gXg^{-1},Q',\a') \text{ with } \\
Q' &=gQg^{-1}-[gXg^{-1},P]+ \frac12[g\epsilon g^{-1},[gXg^{-1},g\epsilon
g^{-1}]
]
\\  &\,\,+\frac12[gXg^{-1},
[g\epsilon g^{-1},\epsilon]] -\frac12[\epsilon,[gXg^{-1},\epsilon]]\\
\a' &=\a-\frac{1}{32}\TR\epsilon[[\epsilon,gXg^{-1}],[\epsilon,P]]-
\frac{1}{8}\TR Q(g^{-1}\epsilon g- \epsilon)\\
&\,\,-\frac{1}{16}\TR X(8 g^{-1}\epsilon
g- 8\epsilon+ [\epsilon,[\epsilon,g^{-1}\epsilon g]+[g^{-1}\epsilon g,
[\epsilon,g^{-1}\epsilon g]]).    \tag3.11\endalign$$

\vskip 0.3in
\bf 4. Vacuum expectation values of operator products\rm

\vskip 0.3in
To start with I shall reformulate some of the results of [PS] on
representations
of $\GL 1$ in a language suited for a generalization to $\GL 2.$

Let $w\in St_1^{(k)}.$ We define a holomorphic section $\psi_w$ of $DET_1^*$ by
$$\psi_w(u)=\text{det} w^*u=\text{det}(w_+^*u_+ + w_-^*u_-),\phantom{aa}
\text{for $u\in St_1^{(k)}$} \tag4.1$$
and $\psi_w(u)=0$ otherwise. We shall think of the basis $w$ as $\Bbb Z\times
\Bbb N$ matrices, the second index labels the different vectors of the basis
and the first index labels the coordinates of the vectors in the standard basis
${e_n}.$ The $w_+$ part of the matrix consists of the rows labelled by
nonnegative coordinate indices and $w_-$ is the lower part consisting of rows
labelled by negative indices. The blocks $w_-,u_-$ are Hilbert-Schmidt matrices
whereas $w_+ -1$ and $u_+ -1$ are of trace-class. It follows that $w^*u$ is
of the form $1+$ a trace-class operator, and the determinant is well-defined.

An inner product in the space of finite linear combinations of the sections
$\psi_w$ is defined by declaring
$$<\psi_w,\psi_w'>=\text{det} (w^*w')\tag4.2$$
when both $w,w'\in St_1^{(k)}$ and the inner product is zero if
the Fredholm indices of $w,w'$ do not coincide.
The Hilbert space completion of this inner product space is denoted by $\Cal
F.$
It can be identified as the fermionic Fock space as follows. Any increasing
sequence $(i_0,i_1,\dots)$ of integers such that lim$(i_{\mu}-\mu)=k$ (where
the
integer $k$ is called the index of the sequence $(i)$) defines a basis vector
in $\Cal F$, $\psi_{(i)}=\psi_w$ with $w=\{e_{i_0},e_{i_1},\dots\}.$ These
vectors are orthonormal with respect to the inner product (4.2). The
interpretation of $\psi_{(i)}$ is that it represents a Fock space state with
holes in the negative energy sea corresponding to one-particle energy levels
labelled by the negative integers in the sequence $(i),$ and filled positive
energy states labelled by the \it missing \rm nonnegative integers in $(i).$

An element $(g,q)\in \GL 1$ acts on the vectors $\psi_w$ by
$$T(g,q)\psi_w=\psi_{g^*w(q^*){-1}}\tag4.3$$
corresponding to the natural action $[T(g,q)\psi](u)=\psi(g^{-1}uq)$ on a
general section of $DET_1^*.$ As already mentioned, in the case $p=1$ we may
choose $q$ to be independent of $W\in Gr_1.$ This means that we can define
$\GL 1$ to be a \it central extension \rm of $GL_1,$ [PS].

An element $W\in Gr_1$ determines the vector $\psi_w$ up to a phase by
selecting
$w$ to be an orthonormal basis of $W.$ A rotation of the basis $w$ by an
unitary
transformation $t\in GL^1$ changes $\psi_w$ by the factor $\text{det}\,t^*.$
Thus the general vacuum expectation values $<W|X|W>:=<\psi_w, X\psi_w>$ of some
operator $X$ in $\Cal F$ depends only on the point $W$ on the base manifold
$Gr_1$ and not on the choice of $w.$

The inner product $<\psi_w,\psi>,$ where $\psi$ is an arbitrary element of
$\Cal F,$ can be computed in the following way. Without an essential
restriction
we may assume that $w$ is orthonormal. Let $\{f_n\}_{n\in\Bbb Z}$ be an
orthonormal basis of $H$ such that $\{f_n\}_{n\geq 0}\in St_1^{(k)}.$ For any
pair $w, w'\in St_1^{(k)}$ we can write
$$\text{det}(w^*w')=\psi_w(w')\tag4.4$$
and therefore
$$<\phi,\psi_w>=\phi(w)\tag4.5$$
for any $\phi\in\Cal F.$ Thus we have:
\proclaim{Proposition 4.6} Let $W\in Gr_1$, $w$ an admissible orthonormal basis
of $W$, and $X\in\gl 1.$ Then
$$<W|X|W>=(X\psi_w)(w).$$ \endproclaim

We shall now go over to the case $p=2$, corresponding to the dimension three of
the physical space. In higher dimensions than one, for any given vacuum state
$|W>$ in the fermionic Fock
space, the vector $X|W>$ does not belong to $\Cal F$ when $X$ is a gauge
current
,
or a product of current components.

If $w\in St_{4/3}$ then the formula  (4.1) still defines a holomorphic section
of $DET_2^*.$ This follows from the generalized H\"older inequalities which
state that $AB$ has a finite trace if $A\in L_p$ and $B\in L_q$ with $\frac1p
+\frac1q=1.$ In the case at hand, $w_-\in L_{4/3}$ and $u_-\in L_4$ and so
$w_-^* u_-\in L_1;$ note that both $w_+ -1$ and $u_+ -1$ are still in $L_1.$

Thanks to the smooth action of $\GL 2$ in $DET^*_2$ the section $X\psi_w$ is a
perfectly well-defined smooth section of
the determinant bundle $DET_p^*$, for any $X\in\gl 2$ or any product of
elements
of $\gl 2,$  and therefore we may evaluate
$$(X\psi_w)(w).\tag4.7$$
We take this expression as the definition of the vacuum expectation
value\newlin
e
$<W|X|W>.$ In order to further motivate this choice, we shall show later that
(4.7) is actually obtained by computing fiberwise the expectation values of
the current algebra in a Fock bundle.

For any finite transformation $g\in GL_2$ we can compute $<H_+|T(g,q)|H_+>$
quite explicitly. The result is
$$<H_+|T(g,q)|H_+>=\text{det}(a\,q(H_+)^{-1}),\tag4.8$$
where $a=a(g)$ is as in (1.3).
In particular, near the unit element in $GL_2$ we may choose $q(H_+)=a$ and
then
the vacuum expectation value at the true vacuum is normalized to one. Of
course,
this does not mean that all vacuum expectation values for elements of $\GL2$
are equal to one, but only for those elements given by the local section
$GL_2\to \GL 2$ above.

The vacuum expectation values of Fock space operators have been determined
earli
er
by Ruijsenaars, [R]. He did not discuss the the abelian extension $\GL 2.$ His
result should be interpreted in the present context as vacuum expectation
values
with respect to some fixed local section $GL_2\to \GL 2.$
Recently Langmann has defined a product for the sesquilinear
forms in Fock space which corresponds to the product of bundle maps discussed
above, [L].

Note that the value of the \it spherical function \rm $\Phi(g,q)=<H_+|T(g,q)|
H_+>$ at $(g,q)$ is equal to the value of the vacuum section $\psi_0$ (which is
defined by the basis $w=w(0):=\{e_0,e_1,\dots\}$), at the point $gwq^{-1}.$
Therefore all the derivatives
$$<X_1X_2\dots X_n>=\frac{d}{dt_1}\dots\frac{d}{dt_n}\Phi(e^{t_1 X_1}\dots
e^{t_n X_n} (g,q))\vert_{(g,q)=1,t_1=\dots=t_n=0},\tag4.9$$
which are the vacuum expectation values of the operator products $X_1\dots
X_n,$
can be written as
$$<X_1\dots X_n>= (X_1\dots X_n\psi_0)(w(0)).\tag4.10$$

The computation of the vacuum expectation values (at the free vacuum $\psi_0=
|H_+>$) is now
completely algebraic. The vacuum $\psi_0$ is annihilated by all the operators
$X\in\bold{gl}_2\subset\gl 2$ with $b(X)=0.$ On the other hand, if $\psi$ is
any section then
$$(X\psi)(w(0))=0 \phantom{aaa}\text{when $c(X)=0$}.\tag4.11$$
This follows from $g^{-1}w(0)q=w(0)$ for any $g$ with $c(g)=0$
and $q=a(g).$ Thus we have the following result:

\proclaim{Theorem 4.12} The vacuum expectation value $<X_1 X_2\dots X_n>$ of
elements $X_i\in\gl 2$ is computed as follows.
Commute all generators $\left(\smallmatrix 0&b\\0&0\endsmallmatrix
\right)$ to the left and all generators of the type $\left(\smallmatrix
a&0\\c&d\endsmallmatrix\right)$ to the right, using the commutation relations
(3.6). Any element on the latter type on the right gives zero when acting on
the vacuum and any element of the former type on the left gives zero when
it hits the vacuum on the left. We are finally left with some function $h$
on the Grassmannian, involving
products of the cocycles $c_2(X,Y;F)$ sandwiched between the vacuum states.
The value of $<X_1\dots X_n>$ is then obtained by evaluating $h$ at the base
point $F=\epsilon.$\endproclaim

\proclaim{Corollary 4.13} $<X>=<XY>=<XYZ>=0$  and
$<XYZV>=$\newline
$\TR\, V_{12}X_{21}Z_{12}Y_{21}+\TR\, X_{21}V_{12}_Y_{21}Z_{12}$
for all $X,Y,Z,V\in\bold{gl}_2\subset\gl 2.$
\endproclaim

Note the difference, as compared to the case of $\gl 1,$  already in the
expectation value $<XY>.$ For $\gl 1$ this is equal to $\TR(X_{21}Y_{12})$
and for $X=Y^*$ gives the norm squared of the vector $X|0>.$ For $\gl 2,$
$<Y^*Y>$ is not really a norm of any vector in the Fock space.

The vacuum expectations are closely related to cyclic cocycles of a Fredholm
module in  Connes' noncommutative geometry, [C]. Let $X_1,\dots,X_k$ be
elements
of $\bold{gl}_p$ with $k\geq 2p.$ Then the trace  \redefine\e{\epsilon}
$$\tau(X_1,X_2,\dots,X_k)=\frac{1}{2^k}\TR \e[\e,X_1][\e,X_2]\dots [\e,X_k]$$
is finite and defines a cyclic cocycle in $\bold{gl}_p.$ It is a simple matter
 to show that
$$\tau(X,Y,Z,V)+\tau(X,V,Z,Y)=<VYZX>-<XZVY>.$$

I want to stress once more that we do not have a true unitary representation of
the Lie algebra $\widehat{\bold{gl}_2}$ but a 'pseudorepresentation'
which allows to compute expectation values of products of generators
compatible with the commutation relations of $\gl 2$ although the generators
are not represented by linear operators in a Fock space. Nevertheless,
the vacuum expectation values can be recovered form a group action in the
Fock bundle, as will be explained below.

The bundle of Fock spaces $\Cal F_W$ parametrized by points $W\in Gr_2$ is
constructed essentially generalizing the construction of the Fock space
$\Cal F=\Cal F_{H_+}$ to arbitrary position on $Gr_2,$ [M1]. Away from the
base point $W=H_+$ there is a modification of the inner product defined by the
formula (4.2), which is related to the fact that in the case $p=2$ the
admissibl
e
basis  $w$ are in general nonunitary. However, in this paper we shall
localize the inner products at the point $H_+$ and therefore we leave it to
the interested reader to check the general case in [M1].
The dual determinant bundle $DET_2^*$ over $Gr_2$ can be thought of as a
one-dimensional complex subbundle of $\Cal F,$ the \it vacuum line bundle. \rm

Given $W\in Gr_2$
the vacuum $Vac(W)$ is spanned by the section $\tilde\phi_w,$ $\tilde\phi_w(u)=
\text{det}
u(w),$ where $w$ is an admissible basis of $W$ and the argument $u\in St^{(0)}
(W)$ and $u(w)$ is the matrix defining the projection of $u$ onto the vectors
$w_i.$ If $u\in St^{(k)}(W)$ with $k\neq 0$ then we set $\tilde\phi_w(u)=0.$

The vacuum vector is well-defined up to a phase (depending on the choice of
$w$). A section of $Vac$ is given by $W\mapsto \phi_W\in\Cal F_W,$
$$\phi_W(u)=\text{det}( w_+)\,\text{det}(u(w)),\tag4.14$$
Note that the values
of $\phi$ do not depend on the choice of $w.$ The second factor in $\phi$ is
invariant under the combined action of $\GL 2$ on the base parameter $w$ and on
the fiber variable $u.$ On the other hand, the first factor is
the highest weight vector in the space of sections $\Gamma(DET_2^*).$
Thus $\phi$ is annihilated by
all $(X,0)\in \widehat{\bold{gl}}_2$ with $b(X)=0.$ Taking into account (4.11)
we get:
\proclaim{Theorem 4.15} For any elements $X_i\in \gl 2,$
$$<\phi, X_1X_2\dots X_n \phi>_{H_+}=<X_1X_2\dots X_n>$$
where $<\cdot,\cdot>_{H_+}$ denotes the fiber inner product at $W=H_+$ in the
Fock bundle over $Gr_2$ and $\phi$ is the vacuum section defined by (4.14).
\endproclaim

In section 5 we shall see how the construction of $\GL 2$ representations is
related to Kirillov's theory of coadjoint orbits. At this point let us make
some preliminary observations in relation to the vacuum expectation value
computations. Let $U_2\subset GL_2$ be the unitary subgroup and $\hat U_2$ the
corresponding extended group; the functions $h$ in (3.5) are then purely
imaginary and $P,Q$ in (3.8) are antihermitean. If $\phi'$ is a section of the
Fock bundle which lies on the
$\hat U_2$ orbit through $\phi,$  $\,\phi'=T(g,q)^{-1}\phi,$ then the natural
de
finition
for the expectation values with respect to the $\phi'$ state is
$$<\phi'|X_1\dots X_n|\phi'>=<\phi|T(g,q)X_1\dots X_n T(g,q)^{-1}|\phi>.
\tag4.16$$
The right-hand side of (4.16) can be computed using the adjoint action of
$\hat U_2$ on its Lie algebra. This is given by the formula (3.11) (for the
larger group $\GL 2.$) Modulo the center, $(g,q)$ can be represented by an
element $(g,P)\in GL'_2.$

Restricted to the subspace $\gl 2$ in the enveloping algebra, the vacuum
expecta
tion
values define an element in the dual space $\gl 2^*.$ Using the notation at the
end of the previous section, a general element in $\gl 2$ is a triple
$(X,Q,\a),
$
where $X\in \bold{gl}_2,\, \a\in \Bbb C,$ and $Q\in M_{2,4/3};$ $Q$ represents
the function $\frac{1}{8}\TR(\epsilon-F)Q$ on the Grassmannian. The vacuum
expectation value with respect to the $\phi$ vacuum is simply $(X,Q,\a)\mapsto
\a.$ By formula (3.11) the vacuum expectation value with respect to $\phi'$ is
the form
$$\align \xi(X,Q,\a)&=<\phi'|(X,Q,\a)|\phi'>=\a-\frac{1}{32}\TR\epsilon[
[\epsilon,gXg^{-1}],[\epsilon,P]]-\frac{1}{8}\TR Q(g^{-1}\epsilon g-\epsilon)\\
&\,\,-\frac{1}{16}\TR X(8 g^{-1}\epsilon
g- 8\epsilon-[\epsilon,[\epsilon,g^{-1}\epsilon g]-[g^{-1}\epsilon g,
[\epsilon,g^{-1}\epsilon g]]).    \tag4.17\endalign$$

The form $\xi\in\gl 2^*$ depends on $(g,P)$ only through $G=g^{-1}\epsilon g$
and $P'=[G,g^{-1}Pg].$ The latter is a hermitean operator (for a unitary $g$)
which
anticommutes with $G.$ The operator $G$ is a point on $Gr_2$ and $P'$ can be
interpreted as a (real) cotangent vector, $P'\in T^*_G Gr_2,$ $P'(u)=\TR uP',$
where $u\in T_G Gr_2.$ Namely, by H\"older inequalities $P'\in M_{2,4/3};$ on
th
e
other hand, the diagonal blocks of $u$ are in $L_2$ whereas the off-diagonal
blocks are in $L_4.$ Again by H\"older inequalities the trace $\TR uP'$
converge
s
for all $u$ if and only if $P'\in M_{2,4/3}.$ Thus the vacua $\phi'=T(g,q)^{-1}
\phi$ (and the linear forms $\xi$) are parametrized by points in $T^* Gr_2.$

\vskip 0.2in
\it More general matrix elements\rm

\vskip 0.2in
Let  $\bold{gl}_0 \subset\bold{gl}_2$ be the subalgebra consisting
of matrices $g$ in (1.3) with finite rank off-diagonal blocks. Denote
$$\b(X;F)=\frac{1}{16}\TR[\epsilon,F][\epsilon,X] \text{ with } X\in
\bold{gl}_0, F\in Gr_2 \tag4.18$$
By a simple computation,
$$\b([X,Y];F)-\Cal L_X\b(Y;F)+\Cal L_Y\b(X;F)=-\frac18 \TR F[[\epsilon,X],
[\epsilon,Y]]\tag4.19$$
and therefore the commutation relations for the generators $\eta(X)=(X,\b(X;
\cdot))$ (in the parametrization (3.4)-(3.5)) are
$$[\eta(X),\eta(Y)]=(\eta([X,Y]), \frac18\TR\epsilon[[\epsilon,X],[\epsilon,
Y]]),\tag4.20$$
It follows that the generators $\eta(X)$, for $X\in\bold{gl}_1$, can be used to
span a representation space for a highest weight representation of
$\gl 1$ by acting by polynomials of the generators on the vacuum vector.
Note that $\b(X;\epsilon)=0,$ thus the highest weight vector for $\gl 2$
is a highest weight vector for $\gl 1.$

The mixed commutation relations between the $X$ and $\eta(X)$ generators are
$$[(X,0),\eta(Y)]=([X, Y], \frac{1}{16}\TR [[F,X],\epsilon][\epsilon,Y])
\tag4.21$$
where the cocycle converges for $X\in \bold{gl}_2, Y\in\gl 0.$
Note that the cocycle converges actually for a slightly larger class of
operators, namely for those $Y$ with $L_{4/3}$ off-diagonal blocks.

The subspace $\Cal F_0$ of the Fock space consisting of finite linear
combinations of vectors of finite energy and finite particle number is
dense in $\Cal F.$ In the charge zero sector $\Cal F^{(0)}$ a Fock basis can
be constructed of vectors in $\Cal F_0$
by applying the Weyl basis generators $\eta(E_{ij})$ (the only nonzero matrix
element of $E_{ij}$, which is equal to 1, is in the $(ij)$ position) to the
vacuum vector.
The inner product of a pair of states $q|0>$ and $p|0>$ is $<0|p^*q|0>$
where the antiautomorphism of the enveloping algebra of $\gl 1$ is determined
by the hermitean conjugation of the Weyl basis generators, $\eta(E_{ij})^*=\eta
(E_{ji}).$
Thus for each polynomial $p(X_1,\dots,X_n)$ in the generators $X_i\in \bold
{gl}_2$ there is a sesquilinear form
$$(\psi,\psi')\mapsto <\psi|p(X)|\psi'>,\tag4.22$$
with domain $\Cal F^{(0)}_0\times\Cal F^{(0)}_0.$ Again, the value of the form
f
or
given pair of vectors $\psi=q(X)|0>,\psi'=q'(X)|0>$ is evaluated by shifting
the
raising generators to the right and lowering generators to the left and finally
evaluating the cocycles at the base point $F=\epsilon,$ Theorem 4.12.

This method works also in any highest weight representation of $\gl 1,$
since it uses only the commutation relations and the existence of a highest
weight vector. However, in the special case of the Fock representation there
is a simpler way to compute matrix elements of products of generators.

Consider the associative algebra $\Cal B$ which is the tensor product of the
enveloping algebra of $\gl 2$ and the CAR algebra based on $H$, modulo
the additional relations
$$[X,a^*(u)]=a^*(Xu), \phantom{A}[X,a(u)]= -a(X^*u)\tag4.23$$
$$[h,a^*(u)]=0=[h,a(u)]$$
where $u,v\in H$, $X\in \bold{gl}_2,$ $h$ is a function on $Gr_2$, and $Xu$
means the defining action in the one-particle space.

By using the commutation relations an arbitrary element of $\Cal B$ can be
written as a linear combination of terms of the type
$$R=X_1\dots X_n h P Y_1\dots Y_k, \tag4.24$$
where $P$ is an element of the CAR algebra,  $h$ is a function on
$Gr_2$, the generators $X_i\in \bold{gl}_2$ have nonzero elements only in $b$-
position, and the $b$-blocks of the $Y_i$'s are zero. The rule for computing
the vacuum expectation value of (4.22) is now the following: $<R>=0$ if $k>0$
or $n>0$. Otherwise $<R>$ is obtained by evaluating the function $h$ at the
base point $F=\epsilon$ and then computing $<P>$ in the usual way using the
basic canonical anticommutation relations.

\vskip 0.2in
\it Matrix elements in an arbitrary highest weight representation\rm

\vskip 0.2in
We can consider also more general highest weight pseudorepresentations.
Let $\l=\{\l_n\}_{n\in\Bbb Z}$ be a sequence of integers, with $\l_n\geq \l_m$
for $n>m,$  such that
$\l_n\to k$ when $n\to\infty$ and $\l_n\to 0$ when $n\to -\infty,$ where
the integer $0\leq k$ is the \it level \rm of the representation.
We shall first consider a (true) representation of $\hat{\bold{gl}}_1,$ [KP].
The representation in a vector space $\Cal F(\l)$ is characterized by a
cyclic vector $|\l>$ such that
$$E_{ij}|\l>=0 \text{ for } i<j \text{ and }  E_{ii}|\l>=\cases
(\lambda_i-k)|\lambda>, \text{for } i\geq 0\\
\lambda_i|\lambda_i> \text{ for } i<0. \endcases \tag4.25$$

The irreducible representation of $\gl 1$ in the charge zero sector of the
fermi
onic
Fock space $\Cal F$ is obtained in the case $k=1,$ $\l_i=1$ for
$i\geq 0$ and $\l_i=0$ for $i<0.$

To each $X\in \gl 2$ we can now associate a sesquilinear form on $\Cal F(\l).$
Let $\gm_+\subset \gl 2$ be the subalgebra of lower triangular matrices $g$,
$g_{ij}=0$ for $i\geq j,$ $\gm_-$ the subalgebra of upper triangular matrices,
$\hm$ the Cartan subalgebra of diagonal matrices. (Note the ordering of matrix
elements; the elements with bigger indices are before those with smaller
indices when reading from top to bottom or from left to right.) Any element of
the enveloping algebra of $\gl 2$ can be written as a sum of products
$$v=u_- z h u_+$$
where $u_{\pm}\in \gm_{\pm},$ $h\in\hm,$ and $z\in Map(Gr_2,\Bbb C).$ The
vacuum expectation value of an arbitrary element in the enveloping algebra
is then evaluated by a straightforward generalization of the previous rules:
An element of $\gm_+$ acting on the vacuum on the right gives zero as well as
an element of $\gm_-$ hitting the vacuum on the left; the vacuum is an
eigenvect
or
of $h\in \hm,$ by (4.25). After these identifications one is left with
expressio
ns
$<z>$ which are evaluated as $<z>=z(\epsilon).$

\vskip 0.3in
\bf 5. Interpretation in terms of coadjoint orbits\rm

\vskip 0.3in
Let $G$ be a Lie group and $\gm$ its Lie algebra. The group $G$ acts in the
dual space $\gm^*$ in a natural way,
$$(g\cdot \xi)(x)=\xi(Ad_{g^{-1}}(x)), \text{ where $x\in\gm$, $\xi\in\gm^*$}
\tag5.1$$
and $Ad_g$ denotes the adjoint action of the group on its Lie algebra.

Each $G$ orbit in $\gm^*$ is a symplectic manifold. The symplectic form $\o$
is given by
$$\o_{\xi}(u,v)=\xi([u,v]),\tag5.2$$
where tangent vectors to the orbit through the point $\xi$ are represented by
elements $u,v\in\gm$ using the Lie algebra action on $\gm^*$ (defined by
left invariant vector fields, infinitesimal version of (5.1)).

If $\gm$ is finite-dimensional, the dimension of the orbit $M=G\xi$ is even,
say
$2n$, and the form $\o^n$ defines a volume on the orbit. An orbit $M$ is \it
quantizable \rm if the symplectic form $\o$ is integral, meaning that the
integral of the 2-form $\o$ over a compact submanifold (without boundary)
is always  an integer. Suppose now that $M$ is quantizable. Then
there is a complex
line bundle $E$ over $M$ with a hermitean metric and a connection $\Gamma$ such
that the curvature is equal to $\o.$ In the coadjoint
method of constructing group representations one first defines a Hilbert space
$V=L^2(E,\o^n)$ of square-integrable sections of the line bundle $E,$ [K].

There is an extension $\hat G$ of $G$ which acts in the total space of the
bundle $E,$ lifting the natural $G$ action on the base and preserving the
hermitean metric. When $G$ is semisimple, $\hat G$ is the universal covering
group of $G.$ This action defines an unitary representation of $\hat G$ in
$V.$ In general, the representation is reducible. Sometimes irreducible
representations can be obtained quite naturally if the orbit $M$ has some
additional structure. A famous example is the case when $G$ is compact and
semisimple
and the orbit has maximal dimension (topologically, $M=G/T$ where $T$ is a
maximal torus). In this case the orbit is a complex manifold, the line bundle
$E$ is holomorphic and the space of \it holomorphic sections \rm of $E$
carries an irreducible unitary representations of $G;$ this is the Borel-Bott-
Weyl theorem. In addition, the theorem tells us that all unitary irreducible
representations can be obtained in this way.

The Borel-Bott-Weyl theorem has been extended to the case of an affine
Kac-Moody
group by Pressley and Segal, [PS]. However, they define the inner product in
the
vector space $V$ in an indirect manner. The reason is that in infinite
dimension
s
the meaning of $\o^n$ becomes problematic. In fact, up to now no suitable
measur
e
on the coadjoint orbits of an affine Kac-Moody group is known which would
produce the unitary highest-weight representations.

Another infinite-dimensional example where the coadjoint orbit method has been
succesful (and this is the case we want to generalize) is $G=\hat U_1,$ the
grou
p
of unitary elements in $\GL 1.$ The Lie
algebra, as a vector space, is now $\hat{\bold{u}}_1=\bold{u}_1\oplus i\RM.$
Consider the coadjoint orbit $M$ through the point $\xi,$
$$\xi(x,\a)=-i\a  \text{ with $x\in\bold{u}_1$ and $\a\in i\RM.$}\tag5.3$$
The stability group at $\xi$ is $D\times S^1,$ where $D\subset U_1$
consists of operators with zero off-diagonal blocks. The orbit $M=\hat U_1/
(D\times S^1)$ is the Grassmannian $Gr_1.$

Let us compute the symplectic form $\o$ more explicitly. Let $F\in Gr_1$ and
$u,v\in T_F Gr_1.$ The tangent space at $F$ consists of hermitean
Hilbert-Schmidt operators which anticommute with $F.$ Choose $X,Y\in
\bold{u}_1$
such that $u=[F,X], v=[F,Y].$
This is always possible since $U_1$ acts transitively on $Gr_1.$ The linear
form
$\overline{F}$ on $\hat{\bold u}_1$ corresponding to $F=g^{-1}\epsilon g$ is
$$\overline{F}(x,\a)= \xi(Ad_g(x,\a))=\xi(x',\a-\frac12\TR(F-\epsilon)x)
=-i\a+\frac{i}{2}\TR (F-\epsilon)x,\tag5.4$$
where $x'$ is some element of $\bold u_1.$ The form $\o_F(u,v)$ is equal to the
value of $F$ at $[(X,0),(Y,0)]$ and thus
$$\o_F(u,v)= -\frac{i}{8}\TR\epsilon[[\epsilon,X],[\epsilon,Y]]+
\frac{i}{2}\TR (F-\epsilon)[X,Y]= \frac{-i}{8}\TR F[u,v].\tag5.5$$
Since $u,v$ anticommute with $F,$ this is equal to $-\frac{i}{4} \TR Fuv.$

The form $\o$ is integral. There is a unique (up to equivalence) complex
line bundle over the coadjoint orbit $Gr_1$ with curvature $2\pi i\o.$ In fact,
this line bundle is the dual determinant bundle $DET_1^*.$ There is a
quasi-invariant measure on the base $Gr_1$ and the square-integrable
holomorphic sections carry an irreducible representation of $\GL 1,$ [P2].

Let us denote by
$\L$ the set of  pairs $(\l,k)$ where $\l=(\l_i)$ is a sequence of integers
labelled by $i\in\Bbb Z,$ $k$ a nonnegative integer, such that
$\l_i\mapsto k$
when $i\mapsto +\infty$, $\l_i\mapsto 0$ when $i\mapsto -\infty,$ and
$\l_i\geq \l_j$ for $i>j.$ For a given $\l$ denote by $i_0$ the smallest
integer such that $\l_{i_0}=k.$

More general orbits in $\gl 1^*$ can now be formed by fixing $(\l,k)\in\L$
and  setting $\xi_{\l,k}(0,\a)=-ik\a,\,\xi_{\l,k}(e_{ii},0)=
\l_i-k=0,$ for $i\geq i_0,$ $\xi_{\l,k}(e_{ii},0)=\l_i$ for $i<i_0,$ and
$\xi_{\l,k}(e_{ij},0)=0$ for $i\neq j,$ where $i,j\in\Bbb Z.$ By restriction,
we
have also a form $\xi_{\l,k}:\hat{\bold u}_1\to\Bbb C.$
We consider a coadjoint orbit $G(\l,k)$ through the point $\xi_{\l,k}.$

The stability group $G_{\l,k}$ at $\xi_{\l,k}$ consists of the center
$S^1$ and of the block diagonal unitary matrices commuting with the diagonal
matrix $diag(\l_i).$ The quotient $M_{\l,k}=\hat U_1/G_{\l,k}$ is an
infinite-dimensional flag manifold; in the special case when $\l_i=1$ for
$i\geq 0$ and $\l_i=0$ for $i<0$ we obtain the Grassmannian $Gr_1.$

The manifold $M_{\l,k}$ has a natural complex structure. It can be written
as a quotient of two complex groups, $M_{\l,k}=GL_1/K_{\l},$ where $K_{\l}$
consists of matrices $g$ in $GL_1$ such that $g_{ij}=0$ for $i\geq j+n_j,$
where
$n_j$ is the largest index $n$ for which $\l_n< \l_j,$ see the picture below.

\vskip 2.5in

For a given weight $\l$ we shall denote the blocks in the stability group $G_{
\l,k}$ by $D_{\mu}$, $\mu=0,1,2,\dots, N,$ where the index $\mu=0$ corresponds
to the
infinite block of rows and columns labelled by $i$ with $\l_i=k,$ and in the
other end $\mu=N$ corresponds to the infinite block of rows and columns
labelled by $i$ with $\l_i=0.$
The blocks in between are all of finite size. We denote by $\l(\mu)$ the value
of the components $\l_i$ in the $\mu:th$ block.

In the following discussion it is convenient to make a finite "vacuum
subtraction." Let $\epsilon_0$ be a hermitean operator in $H$ such that
$\epsilon_0^2=1$ and the difference $\epsilon-\epsilon_0$ has finite rank.
Then the difference
$$c-c_0=\frac18\TR\epsilon[[\epsilon,X],[\epsilon,Y]]-\frac18
\TR\epsilon_0[[\epsilon_0,X],[\epsilon_0,Y]]$$
of cocycles is a coboundary of the 1-cochain $\b(X)=-\frac12\TR(\epsilon-
\epsilon_0)X,$ $(c-c_0)(X,Y)=\b([X,Y]).$ We shall now choose $\epsilon_0$
to be the diagonal matrix with one's at the positions $(ii)$ for $i\geq i_0$
and minus one's at the positions $(ii)$ for $i<i_0.$ The central extension
$\gl 1$ of $\bold{gl}_1$ will be defined now with respect to $\epsilon_0$
instead of $\epsilon.$

There is a complex line bundle $DET_1^*(\l,k)$ over $M_{\l,k}$ defined as
follows. The sections of $DET_1^*(\l,k)$ are complex valued functions $\psi$
on the group $\hat U_1$ such that
$$\psi(gb)=\psi(g)\cdot \text{det}(D_0 q^{-1})\prod_{\mu=1}^{N-1}
(\text{det}D_{\mu})^{\l(\mu)},\tag5.6$$
where $b=(D,q),$ $D$ is a block diagonal matrix and $q$ is a matrix of the same
size as $D_0$ such that $D_0q^{-1}$ has a determinant.  Since left and right
multiplications in a group commute, the space of functions with the
characterist
ics
(5.6) is invariant under the left action, thus defining a representation of
$\hat U_1$ in the space of sections of $DET_1^*(\l,k).$ Since the
multiplication
in $\widehat{GL}_1$ is holomorphic, the space $\G_{\l,k}$ of holomorphic
sections carries a representation of $\hat U_1.$

There is a highest weight vector $\psi_{\l,k}$ in $\G_{\l,k},$ which is defined
by
$$\psi(g,q)=\prod_{\mu=0}^{N-1} [\text{det}(C_{\mu})]^
{\l(\mu)-\l(\mu+1)},\tag5.7$$
where $C_{\mu}$ is the infinite matrix obtained from $gq^{-1}$ by taking the
rows and columns labelled by indices of the blocks $D_0,D_1,\dots,D_{\mu}$ and
$q$ is thought of as an element of $U_1$ by completing $q_{ii}=1$ for $i<i_0,$
$q_{ij}=0$ for $i\neq j$ and $i<i_0$ or $j<i_0.$
By elementary properties of determinants, the highest weight vector has the
properties $E_{ij}\psi_{\l,k}=0$ for $i<j,$ $E_{ii}\psi_{\l,k}=0$
for $i\geq i_0,$ and $E_{ii}\psi_{\l,k}=\l_i\psi_{\l,k}$ for $i<i_0.$

The construction above of representations of $\hat U_1$  extends to the
larger group $\hat U_2$ with the following modifications. First,  we should
keep
in mind that the operator $q$ occurring in the formulas is not a constant but
a function on the Grassmannian $Gr_2$ (with base point $\epsilon_0$).
For this reason the group action on $DET^*_2(\l,k)$ is not holomorphic.
The construction of the highest weight vector is the same as in the case
of $\hat U_1.$

We started from coadjoint orbits of the group $\hat U_1$ and noticed that
the line bundles and group actions extend with the above mentioned
modifications
to the larger group $\hat U_2.$ However, the manifolds $U_2/K_{\l}$ are not
coadjoint orbits of $\hat U_2.$ Because of the larger normal subgroup in the
extension, the coadjoint orbits of $\hat U_2$ have roughly speaking twice the
dimension of the coadjoint orbits of $\hat U_1.$ To illustrate this point we
consider the case of the orbit $Gr_1$ as an example.

Let $\xi\in\gl 2^*$ be the linear form defined by $\xi(X,Q,\a)=-i\a,$ where we
are using the parametrization at the end off section 3, $X\in\bold{gl}_2,\,
Q\in M_{2,4/3},$ and $\a\in\Bbb C.$ That is, $i\xi(X,Q,\a)$ is the vacuum
expectation value of $(X,Q,\a)$ with respect to the free vacuum. Using (3.11)
we conclude that in the coadjoint action by an element $(g,P)^{-1}\in GL'_2$
(we are again dropping the
center of $\GL 2$ since it does not contribute to the (co)adjoint action) the
form $\xi$ is transported to the form $\xi_{g,P},$ given by
$$\align i \xi_{g,P}(X,Q,\a)&=\a-\frac{1}{32}\TR\epsilon[[\epsilon,gXg^{-1}],[
\epsilon,P]]-\frac{1}{8} \TR Q(g^{-1}\epsilon g-\epsilon)\\
&\,\,-\frac{1}{16}\TR X(8 g^{-1}\epsilon
g- 8\epsilon-[\epsilon,[\epsilon,g^{-1}\epsilon g]-[g^{-1}\epsilon g,
[\epsilon,g^{-1}\epsilon g]].    \tag5.8\endalign$$
As we noted in section 4, this form can be interpreted as the vacuum
expectation value of $(X,Q,\a)$ with respect to a vacuum on the $\hat U_2$
orbit through the free vacuum. We also saw that the forms $\xi_{g,P}$ depend
only on $G=g^{-1}\epsilon g$ and $[G, g^{-1}Pg],$ the latter operator being
a vector in $T^*_G Gr_2.$ Thus the coadjoint orbit through $\xi$ can be
identifi
ed
as the cotangent bundle $T^* Gr_2.$

The stability group of $\xi$ in $U'_2$
is the semidirect product of $D$ and the abelian group $A$ of
elements  $(1,P)$ where $P_ \in M_{2,4/3}$ commutes with $\epsilon.$ The
quotient manifold $\hat U_2/(D\ltimes A)$ is the cotangent bundle
$T^* Gr_2.$ The $\hat U_2$ action on $T^* Gr_2$ can be described as follows.
Let $P\in T^*_G Gr_2$ and $(h,Q)\in U'_2.$ Then
$$\align (h,Q)\cdot (G,P)&=(hGh^{-1}, hPh^{-1}+[hGh^{-1},Q]\\  &\,\,
+\frac12[[\epsilon,hGh^{-1}-h\epsilon h^{-1}],hGh^{-1}]-
\frac12 h[[\epsilon,G],G]h^{-1}).\tag5.9  \endalign$$

Let $(\delta G,\delta P)$ be a tangent vector to $T^* Gr_2$ at the base point
$(G,P).$ Then from (5.9) one can reduce that the vector field on $T^*Gr_2$
generated by the element $(\frac12\delta G\,G,
\frac14[\epsilon-G,[\epsilon,\delta G\,G]]+\frac14[G,\delta P])$ in the Lie
alge
bra
$\bold{gl}'_2,$ when evaluated
at $(G,P),$ is equal to $(\delta G,\delta P).$ Using the commutation relations
(3.8)-(3.9) and inserting to the formula (5.8) one obtains, after a bit tedious
but completely straightforward algebra, the following formula for the Kirillov
form on the coadjoint orbit $T^* Gr_2:$

$$\align \Omega_{(G,P)}((\delta_1 G,\delta_1 P),(\delta_2 G,\delta_2 P))&=
\frac{i}{32}\TR G([\delta_2 P,\delta_1 G]-[\delta_1 P,\delta_2 G]) \\ \,\,\,
&-\frac{i}{32}\TR(\epsilon-G)[\delta_1 G,\delta_2 G].\tag5.10\endalign$$
The first term on the right is the canonical 2-form on a cotangent bundle
whereas the second term (which depends only on data on the Grassmannian) is
a generator of the second cohomology on $Gr_2,$ [Q].

To each element $X\in\hat u_2$ (or by complexification of the Lie algebra, a
complex function $f_X$ for any $X\in\gl 2$) one can associate the Hamiltonian
function $f_X:T^*Gr_2\to\RM,$ defined by
$$f_X(G,P)=\xi_{G,P}(X).\tag5.11$$
It satisfies
$$\Cal L_X\cdot f_Y=f_{[X,Y]}=\O(\Cal L_X,\Cal L_Y)\tag5.12$$
where we have denoted by $\Cal L_X$ the vector field on the coadjoint orbit
$M=T^* Gr_2$ generated by left translations by $X.$ Denote by $\{\cdot,\cdot\}$
the Poisson brackets determined by the symplectic form $\O.$ Because of (5.12),
$$\{f_X,f_Y\}=f_{[X,Y]}.\tag5.13$$

In geometric quantization one
associates an operator $\hat f$ to each element $f\in C^{\infty}(M).$ Choosing
a
complex line bundle $E$ over $M$ with connection $\nabla$ such that the
curvatur
e
of $\nabla$ is equal to $2\pi i\O$ the quantum operator $\hat f$ for any $f\in
C^{\infty}(M)$ acts on sections of $E$ and is given by
$$\hat f=\frac{1}{2\pi}\nabla_{V_f} -i f\tag5.14$$
where $V_f$ is the Hamiltonian vector field corresponding to $f,$ $V'\cdot f=
\O(V',V_f)$ for any vector field $V'.$

The above general construction gives a representation of $\gl 2$ on sections of
$E.$ For any $X\in\gl 2$ one can associate an operator $\hat X$ by
$$\hat X=\frac{1}{2\pi}\nabla_X-if_X\tag5.15,$$
where for brevity we denote $\nabla_X=\nabla_{\Cal L_X},$ the covariant
derivati
ve
along the vector field generated by the Lie algebra action on $M.$
\proclaim{Proposition 5.16} The formula (5.15) defines a faithful
representation
of $\gl 2$ in the space of sections $\Gamma(E).$ \endproclaim
\demo{Proof} The group $U'_2$ acts clearly effectively on $T^* Gr_2$ and
therefore to prove that the representation is faithful we have to check only
that the center of
$\gl 2$ is represented nontrivially. The Hamiltonian function $f_c=f_{(0,0,i)}$
corresponding to the central Lie algebra element $c=(0,0,i)$ is the constant
function $f_c(G,P)=1.$ The corresponding Hamiltonian vector field is then
identically zero and so the quantum operator is the multiplication operator
$$\hat f_c= -if_c=-i.\tag5.17$$      \enddemo

By (5.10) the symplectic form is a sum of two pieces $\O_0$ and $d\theta,$
where the latter is the canonical exact 2-form on a cotangent bundle and the
former is nonvanishing only along directions on the base $Gr_2.$ The pull-back
of $\O_0$ on $Gr_2$ (with respect to the zero section $Gr_2\to T^* Gr_2$) is
the curvature of the determinant bundle $DET^*_2$ divided by $2\pi i.$  Thus
we may think of $E$ as the pull-back of $DET^*_2$ on $T^*Gr_2.$

The natural polarization on the cotangent bundle leads to Schr\"odinger
picture of quantization of the Hamiltonian functions $f_X.$

\proclaim{Theorem 5.18} The space of sections of $E$ which are covariantly
constant along the fibers of $T^*Gr_2$ (and can be identified as sections of
$DET^*_2$) is invariant under the $\gl 2$ action and so gives
a representation of $\gl 2$ in the space $\Gamma(DET^*_2).$\endproclaim
\demo{Proof}
Denoting by $A$ a (local) potential for the curvature $2\pi i\O_0,$
the quantum operators can be written as
$$\hat X=\frac{1}{2\pi}\Cal L_X + \frac{1}{2\pi}A(X)+i\theta(X)-if_X.\tag5.19$$
{}From (5.8) one can check that $\theta(X)-f_X$ is a function of $G$ only for
any
$X\in\gl 2.$ Since the local potential $A$ is also a function only on the base
$Gr_2,$ the operators $\hat X$ are of the type $\frac{1}{2\pi}\Cal L_X + $ a
multiplication operator by a function on $Gr_2.$ The sections $\psi$ constant
along fibers of $T^*Gr_2$ are characterized by $\nabla_{(0,Q,0)}\psi=\Cal
L_{(0,Q,0)}\psi=0,$ for $Q\in M_{2,4/3}.$ By the remark above, this property is
preserved under the action by the operators (5.19).\enddemo

\vskip 0.5in
\bf References \rm

\vskip 0.3in

[A] Araki, H.: Bogoliubov automorphisms and Fock representations of canonical
anticommutation relations. In: \it Contemporary Mathematics, \rm
 American Mathematical Society vol. \bf 62 \rm (1987).

[C] Connes, A.: Noncommutative differential geometry. Publ. Math. IHES \bf 62,
\rm 81 (1986).

[FS] Faddeev, L. and S. Shatasvili, Theor. Math. Phys. \bf 60, \rm 770 (1984).

[JJ] Jackiw, R. and K. Johnson: Anomalies of the axial vector current.
Phys. Rev. \bf 182, \rm 1459 (1969).

[K] Kirillov, A.A.: \it Elements of the theory of representations. \rm
Springer-Verlag, Berlin, Heidelberg, and New York (1976).

[KP] Kac, V., and D. Peterson: Lectures on infinite wedge representations and
MKP hierarchy. Seminaire de Math. Sup. \bf 102, \rm 141, Montreal University
(1986)

[L] Langmann, E.: Fermion and boson current algebras in 3+1 dimensions.
In Proceedings of the Symposium "Topological and Geometrical Methods in
Field Theory" in Turku, ed. by J. Mickelsson and O. Pekonen

[Lu] Lundberg, L.-E.: Quasi-free "second quantization". Commun. Math. Phys.
\bf 50, \rm 103 (1976)

[M1] Mickelsson, J.: Commutator anomaly and the Fock bundle. Commun. Math.
Phys.
\bf 127\rm, 285 (1990); On the Hamiltonian approach to commutator anomalies
in $3+1$ dimensions. Phys. Lett. B \bf 241, \rm 70 (1990).

[M2] Mickelsson, J.: \it Current Algebras and Groups. \rm Plenum Press,
New York and London (1989).

[M3] Mickelsson, J.: Chiral anomalies in even and odd dimensions. Commun.
Math. Phys. \bf 97, \rm 361 (1985).

[MR] Mickelsson, J. and S. Rajeev:  Current algebras in $d+1$ dimensions
and determinant bundles over infinite-dimensional Grassmannians. Commun.
Math. Phys. \bf 116,\rm 365 (1988).

[P1] Pickrell, D.: On the Mickelsson-Faddeev extension and unitary
representations. Commun. Math. Phys. \bf 123, \rm 617 (1989).

[P2] Pickrell, D.: Measures on infinite-dimensional Grassmann manifolds.
J. Funct. Anal. \bf 70, \rm 323 (1987).

[PS] Pressley, A. and G. Segal: \it Loop Groups. \rm Clarendon Press,
Oxford (1986).

[Q] Quillen, D.: Superconnection character forms and the Cayley transform.
Topology \bf 27\rm, 211 (1988).

[R] Ruijsenaars, S.N.M.: On Bogoliubov transformations for systems of
relativistic charged particles. J. Math. Phys. \bf 18, \rm 517 (1977).
Index formulas for generalized Wiener-Hopf operators and boson-fermion
correspondence in $2N$ dimensions. Commun. Math. Phys. \bf 124, \rm 553 (1989).

\enddocument